\documentclass{article}
\usepackage{amsmath,graphicx}
\usepackage[preprint]{spconf}
\usepackage{caption}
\usepackage{stfloats}
\usepackage{booktabs}
\usepackage{enumitem}
\usepackage{tikz}
\usepackage{pgfplots}
\pgfplotsset{compat=newest}

\newcommand{\ra}[1]{\renewcommand{\arraystretch}{#1}}
\graphicspath{ {Images/} }

\renewcommand{\vec}[1]{\mathbf{#1}}
\newcommand\mat[1]{\mathbf{#1}}

\usepackage{lipsum}

\copyrightnotice{\copyright\ IEEE 2016}
\toappear{To appear in {\it Proc.\ ICASSP2016, March 20-25, 2016, Shanghai, China}}

\title{RECURRENT NEURAL NETWORKS FOR POLYPHONIC SOUND EVENT DETECTION\\IN REAL LIFE RECORDINGS}
\name{Giambattista Parascandolo, Heikki Huttunen, Tuomas Virtanen\thanks{Tuomas Virtanen has been funded by the Academy of Finland, project no. 258708. The authors wish to acknowledge CSC – IT Center for Science, Finland, for computational resources.}}
\address{Department of Signal Processing, Tampere University of Technology}
\begin{document}
\ninept
\maketitle
\begin{abstract}

In this paper we present an approach to polyphonic sound event detection in real life recordings based on bi-directional long short term memory (BLSTM) recurrent neural networks (RNNs). A single multilabel BLSTM RNN is trained to map acoustic features of a mixture signal consisting of sounds from multiple classes, to binary activity indicators of each event class. Our method is tested on a large database of real-life recordings, with 61 classes (\emph{e.g.}\ music, car, speech) from 10 different everyday contexts. The proposed method outperforms previous approaches by a large margin, and the results are further improved using data augmentation techniques. Overall, our system reports an average $F1$-score of 65.5\% on 1 second blocks and 64.7\% on single frames, a relative improvement over previous state-of-the-art approach of 6.8\% and 15.1\% respectively.
\end{abstract}

\section{Introduction}
\label{sec:intro}

Sound event detection (SED), also known as acoustic event detection, 
deals with the identification of sound events in audio recordings. The goal is to estimate start and end times of sound events, and to give a label for each event. Applications of SED include for example acoustic surveillance \cite{harma2005automatic}, environmental context detection \cite{chu2009environmental} and automatic audio indexing \cite{xu2008audio}. 

SED in single-source environment is called \textit{monophonic} detection, which has been the major area of research in this field \cite{heittola2013context}. However, in a typical real environment it is uncommon to have only a single sound source emitting at a certain point in time; it is more likely that multiple sound sources are emitting simultaneously, thus resulting in an additive combination of sounds. Due to the presence of multiple and overlapping sounds, this problem is known as \textit{polyphonic} detection, and the goal of such a SED system is to recognize for each sound event its category 
(\emph{e.g.}, music, car, speech), and its beginning and ending. This task is much more challenging than the monophonic detection problem, because the sounds are overlapping and the features extracted from the mixture do not match with features calculated from sounds in isolation. Moreover, the number of sources emitting at any given moment (polyphony) is unknown and potentially large.

Initial approaches to polyphonic SED include traditional methods for speech recognition, such as the use of mel frequency cepstral coefficients (MFCCs) as features, with Gaussian mixture models (GMMs) combined with hidden Markov models (HMMs) \cite{mesaros2010acoustic,heittola2013supervised}. A different type of approach consists of extracting and matching the sounds in the input to templates in a dictionary of sounds. This can be achieved through sound source separation techniques, such as non-negative matrix factorization (NMF) on time-frequency representations of the signals. NMF has been used in \cite{innami2012nmf} and \cite{dessein2013real} to pre-process the signal creating a dictionary from single events, and later in \cite{heittola2013supervised} and \cite{dikmen2013sound} directly on the mixture, without learning from isolated sounds. The work in \cite{dikmen2013sound} was extended in \cite{mesaros2015sound} making learning feasible for long recordings by reducing the dictionary size.

Other approaches are based on spectrogram analysis with image processing techniques, such as the work in \cite{dennis2013overlapping} that studies polyphonic SED using generalized Hough transform over local spectrogram features. 

\begin{figure}[tb]
\begin{minipage}[b]{1.0\linewidth}
  \centering
  \centerline{\includegraphics[width=8.5cm]{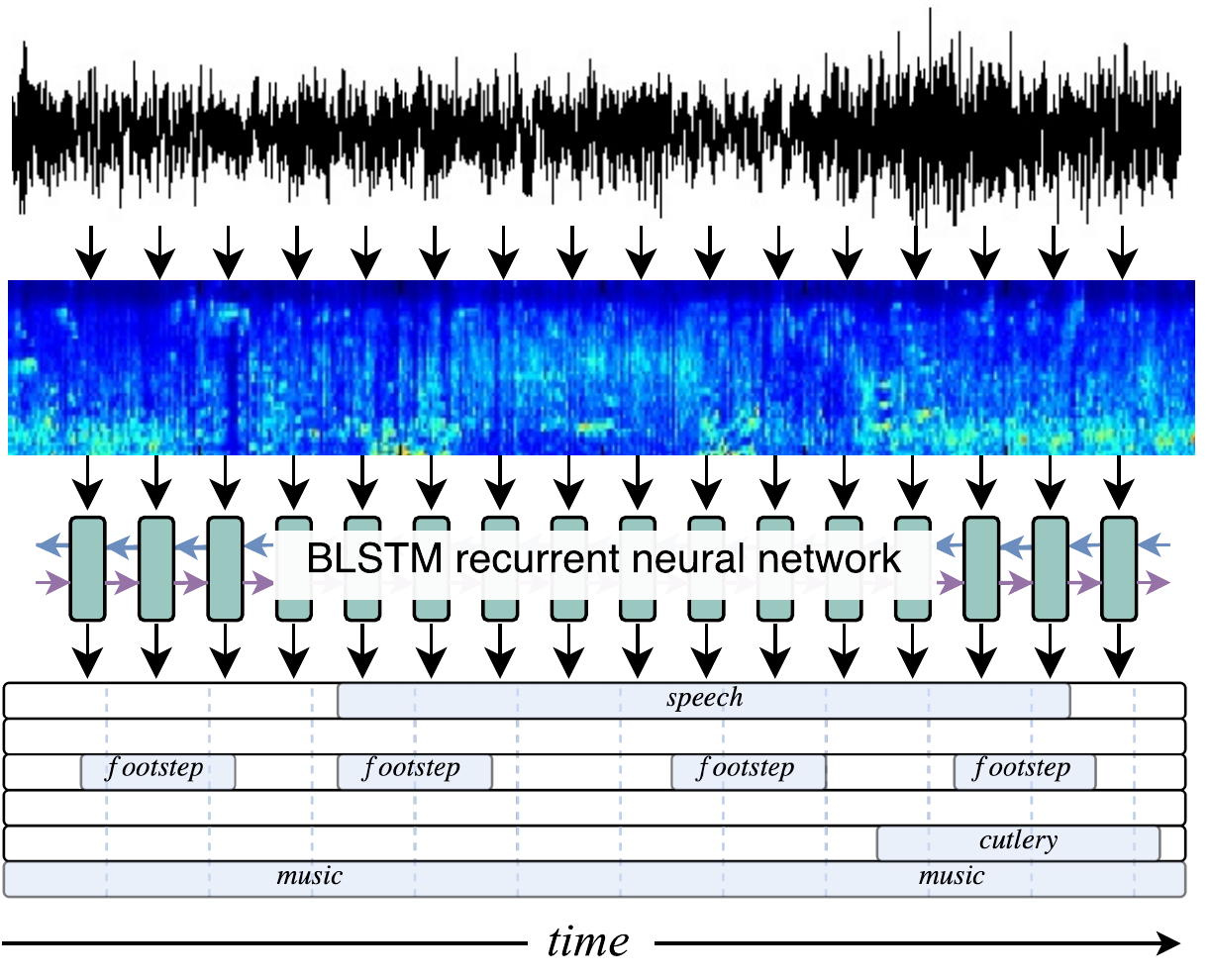}}
\end{minipage}
\caption{Polyphonic sound event detection with BLSTM recurrent neural networks.}
\label{fig:flow_chart}
\end{figure}

More recent approaches based on neural networks have been quite successful. The best results to date in polyphonic SED for real life recordings have been achieved by feedforward neural networks (FNNs), in the form of multilabel time-windowed multi layer perceptrons (MLPs), trained on spectral features of the mixture of sounds \cite{cakir2015ijcnn}, temporally smoothing the outputs for continuity.

Motivated by the good performance shown by the FNN in \cite{cakir2015ijcnn}, we propose to use a multilabel recurrent neural network (RNN) in the form of bi-directional long short-term memory (BLSTM) \cite{hochreiter1997long,graves2005blstm} for polyphonic SED (Fig.\ \ref{fig:flow_chart}). RNNs, contrarily to FNNs, can directly model the sequential information that is naturally present in audio. Their ability to remember past states can avoid the need for tailored postprocessing or smoothing steps. These networks have obtained excellent results on complex audio detection tasks, such as speech recognition \cite{graves2013speech} and onset detection \cite{eyben2010universal} (multiclass), polyphonic piano note transcription \cite{bock2012polyphonic} (multilabel). 

The rest of the paper is structured as follows. Section 2 presents a short introduction to RNNs and long short-term memory (LSTM) blocks. We describe in Section 3 the features used and the proposed approach. Section 4 presents the experimental set-up and results on a database of real life recordings. Finally, we present our conclusions in Section 5.

\section{Recurrent neural networks}
\label{sec:neuralnets}

\subsection{Feedforward neural networks}
In a feedforward neural network (FNN) all observations are processed independently of each other. Due to the lack of context information, FNNs may have difficulties processing sequential inputs such as audio, video and text. A fixed-size (causal or non-causal) window, concatenating the current feature vector with previous (and eventually future) feature vectors, is often used to provide context to the input. This approach however presents substantial shortcomings, such as increased dimensionality (imposing the need for more data, longer training time and larger models), and short fixed context available.

\subsection{Recurrent neural networks}
Introducing feedback connections in a neural network can provide it with past context information. This network architecture is known as recurrent neural network (RNN). In an RNN, information from previous time steps can in principle circulate indefinitely inside the network through the directed cycles, where the hidden layers also act as memory. For a sequence of input vectors $\{\vec{x}_1, ..., \vec{x}_T\}$, a RNN computes a sequence of hidden activations $\{\vec{h}_1, ..., \vec{h}_T\}$ and output vectors $\{\vec{y}_1, ..., \vec{y}_T\}$ as
\begin{align}
	\label{eq:hrnn}
	\mathbf{h}_t &=\mathcal{F}(\mat{W}^{\textnormal{xh}} \mathbf{x}_t + \mat{W}^{\textnormal{hh}} \mathbf{h}_{t-1} + \vec{b}^{\textnormal{h}}) \\
    \mathbf{y}_t &=\mathcal{G}(\mat{W}^{\textnormal{hy}} \mathbf{h}_t + \vec{b}^{\textnormal{y}})
\end{align}
for all timesteps $t = 1, ..., T$, where the matrices $\mat{W}^{\star \star}$ denote the weights connecting two layers, $\vec{b}^{\star}$ are bias terms, and $\mathcal{F}$ and $\mathcal{G}$ activation functions. In case of a deep network, with multiple hidden layers, the input to hidden layer $j$ is the output of the previous hidden layer $j-1$.

When instances from future timesteps are available, also future context can be provided to the network by using bi-directional RNN (BRNN) \cite{schuster1997bidirectional}. In a BRNN each hidden layer is split into two separate layers, one reads the training sequences forwards and the other one backwards. Once fully computed, the activations are then fed to the next layer, giving the network full and symmetrical context for both past and future instances of the input sequence.

\subsection{Long short-term memory}
Standard RNNs, \emph{i.e.}, RNNs with simple recurrent connections in each hidden layer, may be difficult to train. One of the main reasons is the phenomenon called \textit{vanishing gradient problem} \cite{bengio1994learning}, which makes the influence of past inputs decay exponentially over time.

\begin{figure}[htb]
\begin{minipage}[b]{1.0\linewidth}
  \centering
  \centerline{\includegraphics[width=4.5cm]{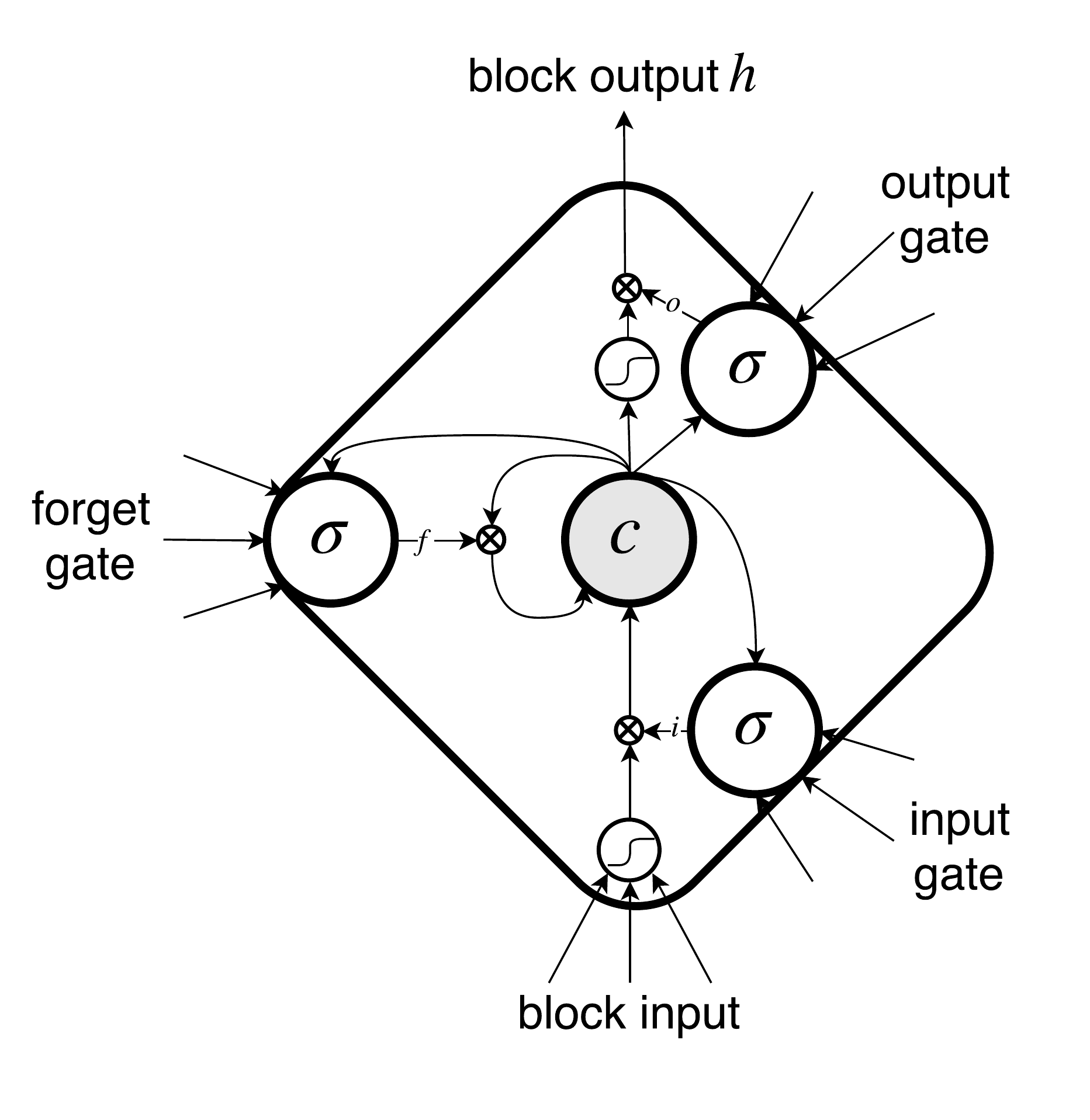}}
\end{minipage}
\caption{An LSTM block.}
\label{fig:lstm}
\end{figure}

The long short-term memory (LSTM) \cite{hochreiter1997long} architecture was proposed as a solution to this problem. The simple neurons with static self-connections, as in a standard RNN, are substituted by units called \textit{LSTM memory blocks} (Fig.\ \ref{fig:lstm}). An LSTM memory block is a subnet that contains one self-connected \textit{memory cell} with its $tanh$ input and output activation functions, and three \textit{gating neurons}---input, forget and output---with their corresponding multiplicative units. Eq.\ \ref{eq:hrnn}, defining the hidden activation $\vec{h}_t$, is substituted by the following set of equations:
\begin{equation}
  \label{eq:LSTM}
\begin{array}{lcl}
\vec{i}_t & = & \sigma(\mat{W}^{\textnormal{xi}} \vec{x}_t + \mat{W}^{\textnormal{hi}} \vec{h}_{t-1} + \mat{W}^{\textnormal{ci}} \vec{c}_{t-1} +  \vec{b}^{\textnormal{i}}) \\
\vec{f}_{t} & = & \sigma(\mat{W}^{\textnormal{xf}} \vec{x}_{t} + \mat{W}^{\textnormal{hf}} \vec{h}_{t-1} + \mat{W}^{\textnormal{cf}} \vec{c}_{t-1} + \vec{b}^{\textnormal{f}}) \\
\vec{c}_{t} & = & \vec{f}_{t} \vec{c}_{t-1} + \vec{i}_{t}\tanh(\mat{W}^{\textnormal{xc}} \vec{x}_{t} + \mat{W}^{\textnormal{hc}} \vec{h}_{t-1} + \vec{b}^{\textnormal{c}}) \\
\vec{o}_{t} & = & \sigma(\mat{W}^{\textnormal{xo}} \vec{x}_t + \mat{W}^{\textnormal{ho}} \vec{h}_{t-1} + \mat{W}^{\textnormal{co}} \vec{c}_{t} + \vec{b}^{\textnormal{o}}) \\
\vec{h}_t & = & \vec{o}_t \tanh(\vec{c}_t)

\end{array}
\end{equation}
where $\vec{c}_{t}, \vec{i}_{t}, \vec{f}_{t}$ and $\vec{o}_{t}$ are respectively the memory cell, input gate, forget gate and output gate activations, $\sigma$ is the logistic function, $\mat{W}^{\star \star}$ are the weight matrices and $\vec{b}^{\star}$ are bias terms.

By analogy, the memory cell $c$ can be compared to a computer memory chip, and the input $i$, forget $f$ and output $o$ gating neurons represent \textit{write}, \textit{reset} and \textit{read} operations respectively.  All gating neurons represent binary switches but use the logistic function---thus outputting in the range $[0, 1]$---to preserve differentiability. Due to the multiplicative units, information can be stored over long time periods inside the cell.

A bidirectional long short-term memory (BLSTM) \cite{graves2005blstm} network is obtained by substituting the simple recurrent neurons in a BRNN with LSTM units. More details about LSTM, BLSTM and training algorithms can be found in \cite{graves2009novel}.

\section{Method}
\label{sec:method}
The proposed system receives as input a raw audio signal, extracts spectral features and then maps them to binary activity indicators of each event class using a BLSTM RNN (Fig.\ \ref{fig:flow_chart}). Each step is described in further detail in this section.


\subsection{Feature extraction}
The input to the system are raw audio signals. To account for different recording conditions, the amplitudes are normalized in each recording to lie in $[-1, 1]$. The signals are split into 50 millisecond frames with 50\% overlap, and we calculate the log magnitudes within the 40 mel bands in each frame. We then normalize each frequency band by subtracting the mean value of each bin over all recordings and imposing unit variance (computing the constants on the training set), a standard procedure when working with neural networks.

For each recording we obtain a long sequence of feature vectors, which is then split into smaller sequences. We split every original sequence at three different scales, \emph{i.e.}, in non-overlapping length 10, length 25, and length 100 sequences (corresponding to lengths of 0.25, 0.62 and 2.5 seconds respectively). This allows the network to more easily identify patterns at different timescales.

Each frame has a target vector $\vec{d}$ associated, whose binary components $d_k$ indicate if a sound event from class $k$ is present or not.

\subsection{Proposed neural network}
We propose the use of multilabel BLSTM RNNs with multiple hidden layers to map the acoustic features to class activity indicator vectors. 
The output layer has logistic activation functions and one neuron for each class. 
We use Gaussian input noise injection and early stopping to reduce overfitting, halting the training if the cost on the validation set does not decrease for 20 epochs. 

The output of the network at time $t$ is a vector $\vec{y}_t \in [0, 1]^L$, where $L$ is the number of classes. Its components $y_k$ can be interpreted as posterior probabilities that each class is active or inactive in frame $\vec{x}_t$. These outputs do not have to sum up to 1, since several classes might be active simultaneously. For this reason, contrarily to most multiclass approaches with neural networks, the outputs are not normalized by computing the softmax. Finally, the continuous outputs are thresholded to obtain binary indicators of class activities for each timestep.

Contrarily to \cite{cakir2015ijcnn}, where the outputs are smoothed over time using a median filter on a 10-frame window, we do not apply any post-processing since the outputs from the RNN are already smooth.

\subsection{Data augmentation}
\label{sec:dataaug}
As an additional measure to reduce overfitting, which easily arises in case the dataset is small compared to the network size, we also augment the training set by simple transformations. All transformations are applied directly to the extracted features in frequency domain.

\begin{itemize}[leftmargin=3.5mm,topsep=0pt]
\item Time stretching: we mimic the process of slightly slowing down or speeding up the recordings. To do this, we stretch the mel spectrogram in time using linear interpolation by factors slightly smaller or bigger than 1;
\item Sub-frame time shifting: we mimic small time shifts of the recordings---at sub-frame scale---linearly interpolating new feature frames in-between existing frames, thus retaining the same frame rate;
\item Blocks mixing: new recordings with equal or higher polyphony can be created by combining different parts of the signals within the same context. In frequency domain we directly achieve a similar result using the mixmax principle \cite{nadas1989speech}, overlapping blocks of the log mel spectrogram two at the time.
\end{itemize}
Similar techniques have been used in \cite{schluterexploring,mcfee2015asoftware}.

The amount of augmentation performed depends on the scarcity of the data available and the difficulty of the task. For the experiments described in Section \ref{sec:eval}---where specified---we expanded the dataset using the aforementioned techniques by approximately 16 times. A 4-fold increase comes from the time stretching (using stretching coefficients of 0.7, 0.85, 1.2, 1.5), 3-fold increase from sub-frame time shifting and 9-fold increase from blocks mixing (mixing 2 blocks at the time, using 20 non-overlapping blocks of equal size for each context). We did not test other amounts or parameters of augmentations. In order to avoid extremely long training times, the augmented data was split in length 25 sequences only.

\begin{table*}[bp]
\centering
\ra{1.1}
\renewcommand\thetable{2}
\caption{Results for each context in the dataset for the FNN in \cite{cakir2015ijcnn} (FNN), and our approach without data augmentation (BLSTM) and with data augmentation (BLSTM+DA).}
\begin{tabular}{@{}lcccccccl@{}}\toprule
& \multicolumn{3}{c}{$F1_{\textnormal{AvgFram}}$} & \phantom{a}& \multicolumn{3}{c}{$F1_{\textnormal{1-sec}}$}\\ \cmidrule{2-4} \cmidrule{6-8}
& $\textnormal{FNN \cite{cakir2015ijcnn}}$ & $\textbf{BLSTM}$ & $\textbf{BLSTM+DA}$ && $\textnormal{FNN \cite{cakir2015ijcnn}}$ & $\textbf{BLSTM}$ & $\textbf{BLSTM+DA}$\\ \midrule
$\textnormal{basketball}$     & 70.2\%& 77.4\% & {\bf 78.5\%}&& 74.7\%& 79.0\% & {\bf 79.9\%}\\
$\textnormal{beach}$ 		  & {\bf 49.7\%}& 46.6\% & 49.6\%&& {\bf 58.1\%}& 48.7\% & 51.5\%\\
$\textnormal{bus}$ 			  & 43.8\%& 45.1\% & {\bf 49.4\%}&& {\bf 52.7\%}& 47.3\% & {\bf 52.7\%}\\
$\textnormal{car}$ 			  & 53.2\%& 67.9\% & {\bf 71.8\%}&& 52.4\%& 66.4\% & {\bf 69.5\%}\\
$\textnormal{hallway}$		  & 47.8\%& {\bf 58.1\%} & 54.8\%&& 55.0\%& {\bf 59.9\%} & 57.1\%\\
$\textnormal{office}$ 		  & 77.4\%& {\bf 79.9\%} & 74.4\%&& 77.7\%& {\bf 79.8\%} & 74.8\%\\ 
$\textnormal{restaurant}$     & 69.8\%& 76.5\% & {\bf 77.8\%}&& 73.7\%& 76.9\% & {\bf 77.7\%}\\
$\textnormal{shop}$ 		  & 51.5\%& {\bf 61.2\%} & 61.1\%&& 57.6\%& 60.9\% & {\bf 61.7\%}\\
$\textnormal{street}$ 		  & 62.6\%& {\bf 65.3\%} & 65.2\%&& 62.9\%& 63.3\% & {\bf 63.9\%}\\
$\textnormal{stadium}$ 		  & 58.2\%& 61.7\% & {\bf 64.3\%}&& 64.9\%& 64.2\% & {\bf 66.2\%}\\
$\textnormal{average}$ 		  & 58.4\%& 64.0\% & \bf{64.7\%} && 63.0\%& 64.6\% & {\bf 65.5\%}\\
\bottomrule
\label{tab:res_big}
\end{tabular}
\end{table*}

\section{Evaluation}
\label{sec:eval}
\subsection{Dataset}
\label{sec:dataset}
\begin{figure}[b]
\centering
\begin{tikzpicture}

\begin{axis}[%
width=2.528in,
height=0.654in,
scale only axis,
xmin=0,
xmax=8,
nodes near coords,
xtick={ 1,  2,  3,  4,  5,  6,  7},
xlabel={Polyphony},
ymin=0,
ymax=43,
ytick={0, 10, 20, 30},
yticklabel=\pgfmathparse{1*\tick}\pgfmathprintnumber{\pgfmathresult}\,\%,
yticklabel style={/pgf/number format/.cd,fixed,precision=2},
ylabel={Percentage of data},
axis background/.style={fill=white}
]
\addplot[ybar,bar width=0.6,draw=black,fill={rgb:red,1;green,2;blue,3},area legend] plot table[row sep=crcr] {%
1	25.7\\
2	29.5\\
3	21.3\\
4	12.5\\
5	7.8\\
6	2.7\\
7	0.3\\
};

\end{axis}
\end{tikzpicture}%
\caption{Distribution of polyphony level across the dataset.}
\label{fig:poly}
\end{figure}
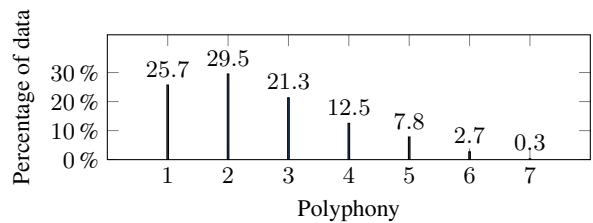

We evaluate the performance of the proposed method on a database consisting of recordings 10 to 30 minutes long, from ten real-life contexts \cite{heittola2010audio}. The contexts are: basketball game, beach, inside a bus, inside a car, hallway, office, restaurant, shop, street and stadium with track and field events. Each context has 8 to 14 recordings, for a total of 103 recordings (1133 minutes). The recordings were acquired with a binaural microphone at 44.1 kHz sampling rate and 24-bit resolution. The stereo signals from the recordings are converted to mono by averaging the two channels into a single one. The sound events were manually annotated within 60 classes, including speech, applause, music, break squeak, keyboard; plus 1 class for rare or unknown events marked as \emph{unknown}, for a total of 61 classes. All the events appear multiple times in the recordings; some of them are present in different contexts, others are context-specific. The average polyphony level---\emph{i.e.}\ the average number of events active simultaneously---is 2.53, the distribution of polyphony levels across all recordings is illustrated in Fig.\ \ref{fig:poly}.

The database was split 
into training, validation and test set (about 60\%, 20\% and 20\% of the data respectively) in a 5-fold manner. All results are presented as averages of the 5-fold cross validation results, with the same train/validation/test partitions used in previous experiments on the same dataset (\cite{mesaros2015sound,cakir2015ijcnn}). The hyperparameters of the network, \emph{e.g.}\ the number and size of hidden layers, learning rate, etc., were chosen based on validation results of the first fold.

\subsection{Neural networks experiments}
The network has an input layer with 40 units, each reading one component of the feature frames, and 4 hidden layers with 200 LSTM units each (100 reading the sequence forwards, 100 backwards). We train one network with the original data only, which is the same used in previous works, and one using the data augmentation techniques reported in Section \ref{sec:dataaug} to further reduce overfitting. To compare the performance with standard LSTM layers, we also train a similar network architecture without bidirectional units on the same dataset without augmentation.

The network is initialised with uniformly distributed weights in $[-0.1,0.1]$ and trained using root mean squared error as a cost function. Training is done by back propagation through time (BPTT) \cite{werbos1990bptt}. 
The extracted features are presented as sequences clipped from the original data---in sequences of 10, 25 and 100 frames---in randomly ordered minibatches of 600 sequences, in order to allow parallel processing. After a mini-batch is processed the weights are updated using 
RMSProp \cite{tieleman2012lecture}. The network is trained with a learning rate $\eta = 0.005$, decay rate $\rho = 0.9$ 
and Gaussian input noise of 0.2 (hyperparameters chosen based on the validation set of the first fold). At test time we present the feature frames in sequences of 100 frames, and threshold the outputs with a fixed threshold of 0.5, \emph{i.e.},\ we mark an event $k$ as active if ${y}_k \geq 0.5$, otherwise inactive.

For each experiment and each fold we train 5 networks with different random initialisations, select the one that has the highest performance on the validation set and then use it to compute the results on the test data. The networks were trained on a GPU (Tesla K40t), with the open-source toolkit Currennt \cite{weninger2015introducing} modified to use RMSprop.

\subsection{Metrics}
To evaluate the performance of the system we compute $F1$-score for each context in two ways: average of framewise $F1$-score ($F1_{\textnormal{AvgFram}}$) and average of $F1$-score in non-overlapping 1 second blocks ($F1_{\textnormal{1-sec}}$) as proposed in \cite{heittola2013context}, where each target class and prediction is marked as active on the whole block if it is active in at least one frame of the block. The overall scores are computed as the average of the average scores for each context.

\subsection{Results}
In Table \ref{tab:res} we compare the average scores over all contexts for the FNN in \cite{cakir2015ijcnn} to our BLSTM and LSTM networks trained on the same data, and BLSTM network trained with the augmented data. The FNN uses the same features but at each timestep reads a concatenation of 5 input frames (the current frame and the two previous and two following frames). It has two hidden layers with 1600 hidden units each, downsampled to 800 with maxout activations.

\begin{table}[h]
\centering
\ra{1.1}
\renewcommand\thetable{1}
\caption{Overall $F1$ scores, as average of individual contexts scores, for the FNN in \cite{cakir2015ijcnn} (FNN) compared to the proposed LSTM, BLSTM and BLSTM with data augmentation (BLSTM+DA).}
\label{tab:res}
\begin{tabular}{ @{} l *{2}{r} *{4}{c} @{} } 
\toprule
Method & $F1_{\textnormal{AvgFram}}$ & $F1_{\textnormal{1-sec}}$\\
\midrule
FNN \cite{cakir2015ijcnn} & 58.4\% & 63.0\% \\
{\bf LSTM}    & 62.5\% & 63.8\%\\
{\bf BLSTM}    & 64.0\% & 64.6\%\\
{\bf BLSTM+DA} & {\bf 64.7}\% & {\bf 65.5}\%\\
\bottomrule
\end{tabular}
\end{table} 

The BLSTM network achieves better results than the FNN trained on the same data, improving the performance by relative 13.5\% for the average framewise $F1$ and 4.3\% for the 1 second block $F1$. The unidirectional LSTM network does not perform as well as the BLSTM network, but is still better than the FNN. 
The best results are obtained by the BLSTM network trained on the augmented dataset, which improves the performance over the FNN by relative 15.1\% and 6.8\% for the average framewise $F1$ and for the 1 second block $F1$ respectively.

In Table \ref{tab:res_big} we report the results for each context for the FNN in \cite{cakir2015ijcnn} (FNN), our BLSTM trained on the same data (BLSTM) and our BLSTM trained on the augmented data (BLSTM+DA). The results show that the proposed RNN, even without the regularisation from the data augmentation, outperforms the FNN in most of the contexts. 


The $F1$-scores for different polyphony levels are approximately the same, showing that the method is quite robust even when several events are combined.
It is interesting to notice that the RNNs have around 850K parameters each, compared to 1.65M parameters of the FNN trained with the same data. The RNNs make a more efficient and effective use of the parameters, due to the recurrent connections and the deeper structure with smaller layers.

\section{Conclusions}
\label{sec:conclusions}
In this paper we proposed to use multilabel BLSTM recurrent neural networks for polyphonic sound event detection. RNNs can directly encode context information in the hidden layers and can learn the longer patterns present in the data. Data augmentation techniques effectively reduce overfitting, further improving performance. The presented approach outperforms the previous state-of-the-art FNN \cite{cakir2015ijcnn} tested on the same large database of real-life recordings, and has half as many parameters. The average improvement on the whole data set is 15.1\% for the average framewise $F1$ and 6.8\% for the 1 second block $F1$.

Future work will concentrate on finding novel data augmentation techniques. Concerning the model, further studies will develop on attention mechanisms and extending RNNs by coupling them with convolutional neural networks.

\vfill\pagebreak

\bibliographystyle{IEEEbib}
\bibliography{database.bib}

\end{document}